\documentclass[conference]{IEEEtran}

\ifCLASSINFOpdf
	\usepackage[pdftex]{graphicx}
    \graphicspath{{}}
    \DeclareGraphicsExtensions{.pdf}
\else
\fi
\usepackage{graphicx}
\ifCLASSOPTIONcompsoc
  \usepackage[caption=false,font=normalsize,labelfont=sf,textfont=sf]{subfig}
\else
  \usepackage[caption=false,font=footnotesize]{subfig}
\fi

\usepackage{colortbl}
\usepackage{cite}
\usepackage{booktabs}
\usepackage{algorithmic}
\usepackage{algorithm} % NB stated not to use it!

\begin{document}
\title{Under Frequency Load Shedding based on PMU Estimates of Frequency and ROCOF}

\author{
\IEEEauthorblockN{Asja~Dervi{\v s}kadi{\'c}, Yihui~Zuo, Guglielmo~Frigo, Mario~Paolone}
\IEEEauthorblockA{Distributed Electrical Systems Laboratory (DESL)\\
\'Ecole Polytechnique F\'ed\'erale de Lausanne (EPFL)\\
Lausanne, Switzerland\\
Email: asja.derviskadic@epfl.ch
}}

\maketitle

\begin{abstract}
The paper describes a distributed under-frequency load shedding and load restoration scheme, that exploits frequency and Rate-of-Change-of-Frequency (ROCOF) measurements produced by Phasor Measurement Units (PMUs) as detectors of a large contingency. 
The paper further discusses the appropriateness of using the synchrophasor model for estimating ROCOF during fast transients following a severe generation-load imbalance. 
The performance of the proposed relaying scheme is compared with a frequency-only based strategy, by means of a real-time digital simulator implementing the time-domain full-replica model of the IEEE 39 bus system. 
\end{abstract}


\IEEEpeerreviewmaketitle

\section{Introduction}
{As is well known, the} under-Frequency (UF) Load Shedding (LS) is a technique that minimizes the risk of uncontrolled system separation, loss of generation, or shutdown~\cite{stdC37.117-2007}. Typically, UFLS schemes rely on updated measurements {of average or local} system frequency: 
once a given threshold is exceeded, dedicated relays automatically trip part of the loads in order to preserve grid interconnections and generation capability. If a sufficient amount of loads is shed, the system {load} can be smoothly and rapidly restored. 
Traditional approaches determine the amount of shed loads based on frequency only~\cite{UFLS_AUPEC2011,LS_TPWRS2013}, but recent literature has considered the adoption of more responsive and effective relays based on the Rate-of-Change-of-Frequency (ROCOF)~\cite{ROCOFUFLS_GM2010,ROCOFUFLS_TPWRS2017}.

Compared to frequency-based solutions (f-LS), ROCOF-based LS (ROCOF-LS) guarantees two major benefits. 
Since ROCOF is defined as the frequency first time-derivative, it can be seen as a predictive filter. 
As soon as the system frequency starts decreasing, ROCOF accounts for the variation polarity and velocity. 
By applying a threshold on ROCOF estimated values, it is possible to promptly detect critical conditions, even before frequency has fallen below abnormal operation levels, and thus guarantee a faster load restoration (LR). 
For the same reason, it is {intuitive} to expect that ROCOF-based solutions would require a smaller amount of shed loads, corresponding also to a smaller amount of curtailed energy. 

In this context, Phasor Measurement Units (PMUs) might play a decisive role for the development of enhanced control schemes that leverage on frequency and ROCOF measurements~\cite{lavertyLOSS2015,LS_TPWRS2013}, taking advantage of PMUs' high reporting rates, remarkable measurement accuracy {as well as time synchronization~\cite{stdc37.118.1-2011} (this last can be used to define time-aware centralized schemes for LS)}.   

In practice, though, two implementation issues arise. 
From a measurement perspective, PMUs rely on a synchrophasor signal model that does not perfectly fit with time-varying conditions, particularly if transient events occur~\cite{Barchi-etAl2013,Frigo-etAl2017}. In this sense, the reliability and accuracy of PMU-based estimates has to be carefully considered, before determining the most suitable threshold level~\cite{KirkhamRTUCON2016,RoscoeAMPS2017}. 
From a control perspective, ROCOF variability does not follow pre-defined statistical distribution, but depends on the characteristics of the considered power network {and on its state before the contingency}. A preliminary analysis of the synchronous area is recommended to determine the expected behavior of the system frequency in critical conditions~\cite{ROCOFrelay_TPWRD2006}. 

In this paper, we describe a simple yet effective distributed UFLS and LR scheme, that relies on PMU-based estimates of frequency and ROCOF. We assess the performance of the proposed relaying scheme by means of a real-time digital simulator (RTS), where we reproduce the time-domain full-replica model of the IEEE 39 bus system, hosting substantial amount of distributed energy resources~\cite{IEEE39bus}.

The paper is structured as follows.
In Section II we discuss the role of ROCOF measurements in modern power system control schemes. Section III presents the proposed UFLS relaying technique. In Section IV we describe the developed RTS test-bed, including simulated C37.118 class-P PMUs. Section V assesses the technique performance in two different scenarios. Finally, Section VI provides some closing remarks and outlines the future stages of the research activity.

\section{PMU-based Measurement of ROCOF}
PMUs are measurement devices providing estimates of voltage and current synchrophasors, frequency and ROCOF associated to the power signal fundamental component. 
These estimates are updated with high reporting rates (in the order of tens of frames-per-second (fps)) and synchronized with respect to Coordinated Universal Time (UTC). The IEEE Std. C37.118.1~\cite{stdc37.118.1-2011} and its recent amendment~\cite{stdc37.118.1a-2014}, define the PMU requirements in terms of accuracy and latency. 

According to the IEEE Std. C37.118.1, fundamental frequency and ROCOF are defined as the first and second time-derivative of synchrophasor phase angle, respectively. However, most PMUs compute ROCOF as the finite difference between two consecutive frequency estimates. 
%\textcolor{red}{REMOVE : This approach introduces a sort of filtering effect on ROCOF estimation. For this reason, in the presence of rapid changes, the PMU produces a delayed response.}
{This approach, as well-known, amplifies the effects of measurement noise on the accuracy of the estimates}.

With specific reference to class-P PMUs, the IEEE Std. C37.118.1 requires the ROCOF error (RFE) not to exceed 10 mHz/s and 3 Hz/s in static nominal and dynamic conditions, respectively\footnote{These requirements have been significantly relaxed in the recent amendment~\cite{stdc37.118.1a-2014}, in view of a more appropriate definition of frequency and ROCOF measurements}.
Most of recent synchrophasor estimation algorithms outperform these requirements and are able to guarantee RFEs below hundreds mHz/s in any test conditions~\cite{Roscoe2013,Castello-etAl2014,Frigo-etAl2015,classPM}.

The synchrophasor models rely on the assumption that the acquired signal spectrum consists of one or more narrow-band spectral components. 
In real-world scenario, such a model might loose its appropriateness. In particular, during transient events, the definition of frequency and ROCOF associated to the fundamental component still represents an open issue from the metrological point of view~\cite{PhadkeTransient2008}.
However, such investigations are beyond the scope of this paper. At the present stage of the research, we develop and characterize an UFLS and LR scheme, independently from the uncertainty inherent in the employed frequency and ROCOF estimates.

Typically, ROCOF-LS (and other ROCOF-based power system applications) is implemented in dedicated relays that compute ROCOF over 500 ms observation intervals~\cite{ENTSOE_rocof}, whereas class-P PMUs adopt observation intervals not exceeding 80 ms and reporting rates not lower than 50 fps. 
In other words, PMUs compute ROCOF as the fundamental frequency time-derivative over 20 ms. This discrepancy produce different accuracy and responsiveness levels. ROCOF-relays are characterized by much smoother output trends, whereas PMU-based estimates are subject to rapid oscillations. By contrast, PMUs promptly detect transient events, that can be neglected or significantly understated by traditional ROCOF-relays. 

\section{ROCOF-based Load Shedding}
UFLS schemes enable preventing power system collapse and blackout in case the frequency drops below critical values after severe system contingencies. Typically, they exploit dedicated relays that curtail specific shares of loads, every time the system frequency exceeds predefined threshold values~\cite{stdC37.117-2007}.
In this regard, the recent literature proposes several f-LS schemes:
as an example, the European Network of Transmission System Operators for Electricity (ENTSO-E) recommends to shed loads proportionally to the measured frequency drop~\cite{ENTSOE_LS}. 

As regards ROCOF-LS, instead, standard or recommended criteria for a suitable definition of threshold values are still missing. By definition, ROCOF accounts for the instantaneous power balance within the considered grid. Therefore, an accurate ROCOF estimation {with reduced latency} would enable us to design more selective and faster control schemes. However, the deployment of ROCOF-relays is still limited to islanding detection for distribution networks~\cite{islandingdet_TPWRD2016} and loss-of-mains protection of embedded generators~\cite{lavertyLOSS2015}. 

\begin{figure}
	\centering
	\includegraphics[width=0.9\columnwidth]{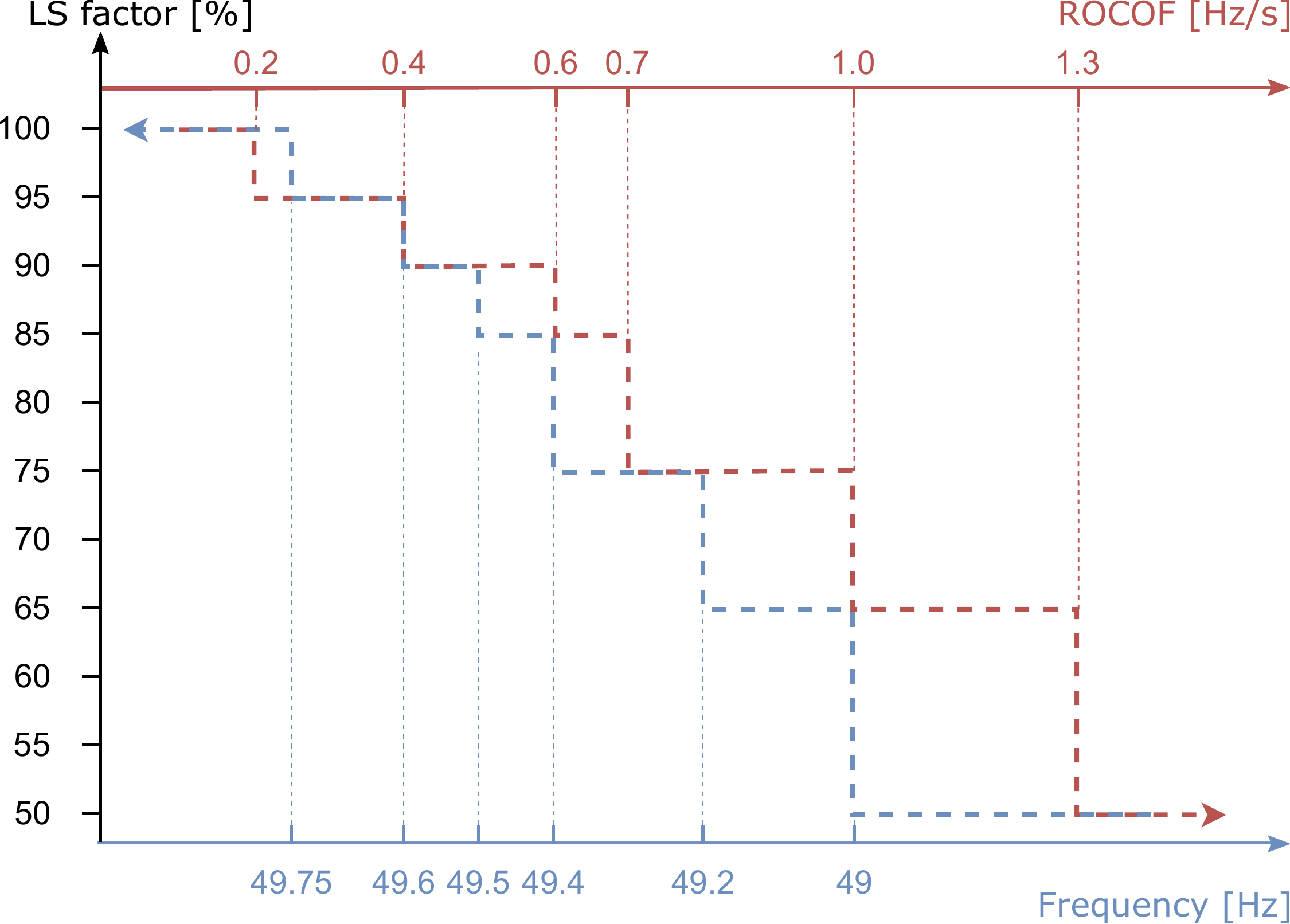}
	\caption{The proposed ROCOF and frequency thresholds for ROCOF-LS A and f-LS respectively.}
	\label{fig:UFLS_thresh}
\end{figure}

Table~\ref{tab:UFLS} 
\begin{table}
	\caption{ROCOF and frequency thresholds for LS and LR}
	\label{tab:UFLS}
	\centering
	\setlength\tabcolsep{4pt}
	\begin{tabular}{ccccccccc}
		\toprule
		LS factor &[\%]  & 100 & 95 & 90 & 85 & 75 & 60 & 50 \\
		\midrule
		ROCOF-LS A &[Hz/s]& & 0.2 & 0.4 & 0.6 & 0.7 & 1 & 1.3 \\
		ROCOF-LS B &[Hz/s]& & 0.2 & 0.3 & 0.4 & 0.5 & 1 & 1.3 \\
		\midrule
		f-LS  &[Hz]& & 48.9 & 48.8 & 48.6 & 48.4 & 48.2 & 48 \\
		\midrule
		f-LR  &[Hz]& 49.75 & 49.6 & 49.5 & 49.4 & 49.2 & 49 & \\
		\toprule
	\end{tabular}
\end{table}
presents the considered threshold values for LS and LR, as a function of the related load percentage.  
For comparison purposes, we implement f-LS, following ENTSO-E's recommendations~\cite{ENTSOE_LS}, as well as ROCOF-LS, tuning sets of thresholds characterized by increasing values of ROCOF (case A and case B). 
By contrast, we implement a single f-LR, referring to the guidelines in~\cite{ENTSOE_LS}. 
Figure~\ref{fig:UFLS_thresh} 
further shows the correlation between ROCOF-LS and f-LR.

The reason why we use ROCOF measurements {to trigger} the LS process only, is that after a specific contingency, ROCOF values experienced by each bus mainly depend on the characteristics of the considered electric grid: in general, for a specific electric grid, the larger the contingency, the larger the ROCOF. Also, a long-lasting negative ROCOF value, unequivocally identifies a load imbalance that must be cleared. 
By contrast, during the network-restoration process, ROCOF values experienced along the grid strongly depend on the adopted restoration actions. Therefore, on the one hand, it is quite difficult to infer all the possible attainable ROCOF values, on the other hand, a long-lasting positive ROCOF value, does not necessarily indicate that the grid has reached a stable status that could handle the connection of further loads. 
The recovery of system frequency towards nominal values, is instead an unequivocal indicator of secure system state. 

Moreover, it is worth pointing out that ROCOF estimates have to be suitably filtered in order to mitigate the instantaneous transients and the longterm-damped oscillations of system frequency after contingency events. In this way, we are able to guarantee more stable and reliable trends, but we also introduce a time-delay that, if uncontrolled, could frustrate the benefits of ROCOF predictive capability. In this respect, we consider a 500 ms time delay for the intervention of ROCOF-relays. 

As the system frequency approaches the nominal value, the restoration of the load that has been shed can start, depending on the ability of the system to serve it. 
Every time a load is restored, an unavoidable decrease in system frequency occurs, that could lead to undesirable LS repetition or oscillation between LS and LR. 
{
Therefore, a LR program typically incorporates a time delay between two subsequent LR steps, to enable the system to stabilize before an additional block of load is picked up. In this respect, we introduce a 5 s time delay after every LR step~\cite{stdC37.117-2007}.
}

It is important to observe that dynamic phenomena following every critical event or control action, cannot be accurately forecast or compensated, since they depend on too many aspects, like the type event, the electrical properties of the grid synchronous area, and the location of metering devices. Therefore, before implementing any ROCOF-based control scheme, a thorough analysis of the power system under investigation could inspire some practical criteria for the definition of threshold values and corresponding shares of loads to be shed. To this end, RTS enables us to reproduce a wide range of realistic operating conditions and thus evaluate the reliability of the proposed ROCOF-based solutions during different critical scenarios.

\section{Simulation model}
\begin{figure}
\centering
\includegraphics[width=1\columnwidth]{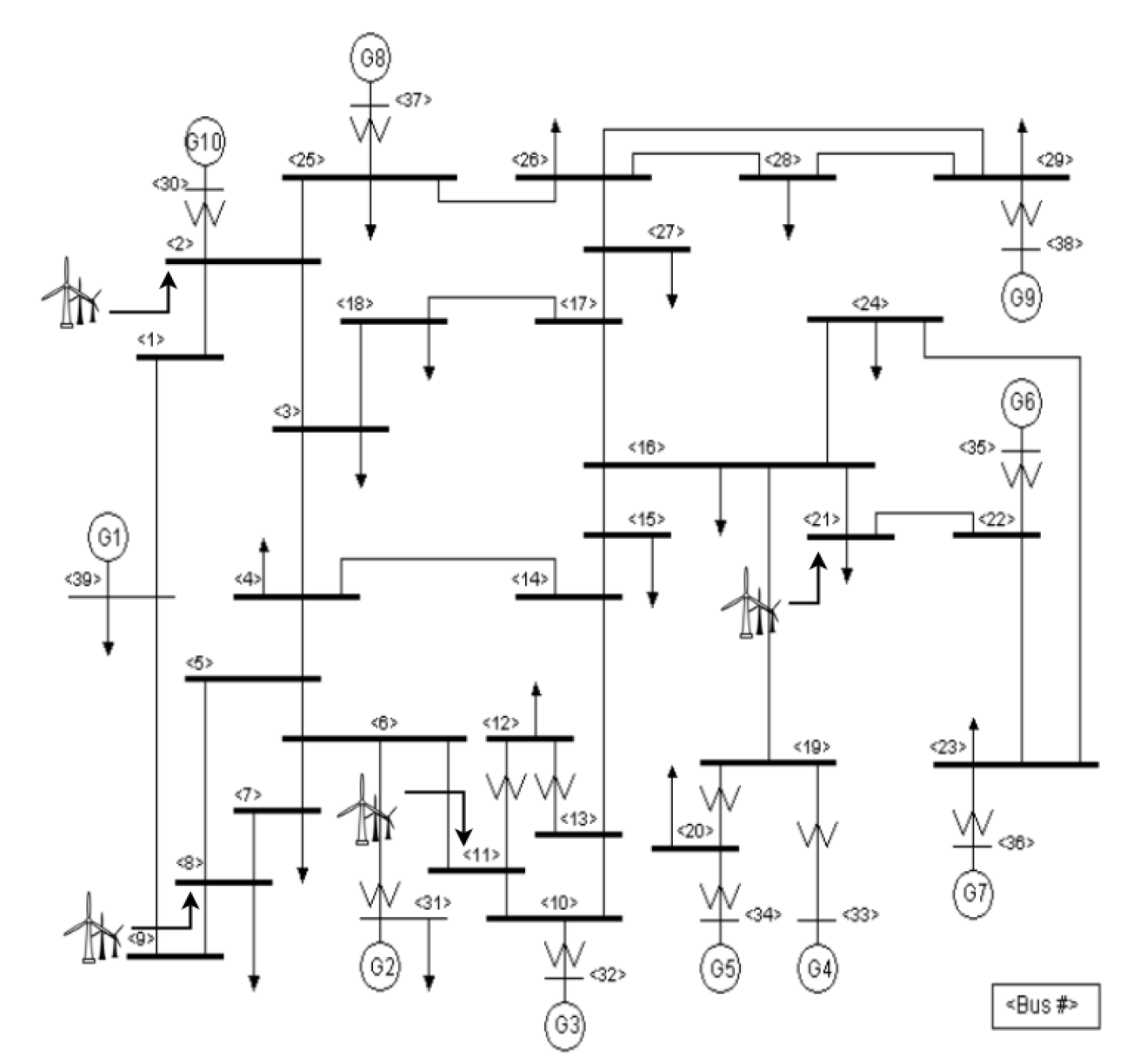}
\caption{Diagram of the modified 39-bus power grid, adapted from~\cite{IEEE39bus}.}
\label{fig:IEEE39bus}
\end{figure}

In order to test the proposed technique, we adopt the Opal-RT eMEGAsim PowerGrid RTS~\cite{OpalRT} to implement a detailed dynamic model of IEEE 39-bus power system, also known as 10-machine New-England Power System~\cite{IEEE39bus}, represented in Fig.~\ref{fig:IEEE39bus}.
In fact, this model represents a widely-employed benchmark for performance evaluation and comparison of several monitoring and control applications. In more detail, the simulated power system has a nominal voltage of 345 kV, and consists of 39 buses, 10 conventional synchronous generators and 19 loads. 
In order to take into account the effects of distributed renewable generation, we modified the benchmark by including 4 wind farms. 
Moreover, in order to emulate realistic load and generation patterns, we used wind and load profiles coming from real measurements. The network as well as the measurement devices are modeled in Simulink and the simulations are run by using the Opal-RT eMEGAsim RTS. This Section presents in detail the main elements included in the model, {further described in~\cite{zuoISGT2018}.}

\subsection{Synchronous generators}
For each conventional generator, Table~\ref{tab:generation} reports plant type, location and nominal capacity.  
\begin{table}
\caption{Generators Type and Capacity}
\label{tab:generation}
\centering
\begin{tabular}{cccc}
\toprule
Unit & Plant Type & Location & Installed Capacity \\
 & & (Bus \#) & [MVA] \\
\toprule
G1 & Thermal & 39 & 3000\\
G2, G3, G4, & Hydro   & 31, 32, 33  & 1000\\
G6, G7, G8, G9 & & 35, 36, 37, 38  & \\
G5 & Hydro   & 34 & 520 \\
\toprule
WF1 & Wind Farm & 2 	& 300\\
WF2 & Wind Farm & 21 	& 150\\
WF3 & Wind Farm & 8 	& 400\\
WF4 & Wind Farm & 11 	& 500\\
\toprule
\end{tabular}
\end{table}
In this regard, the thermal plant has an installed capacity of 3 GVA, whereas two types of hydro-power plant are simulated, characterized by an installed capacity of 1 GVA and 520 MVA, respectively.  
Every power plant model consists in a dynamic model of the prime mover (hydro or thermal), the electric generator (synchronous machine), the speed governor,
% \textcolor{red}{IEEE DCA1-type} 
the exciter and the associated automatic voltage regulator. 
All the synchronous machines are modeled by a sixth-order state-space model available in the SimPowerSystem Simulink toolbox. The inertia coefficients are set equal to those recommended by the 39-buses benchmark. 
It is worth pointing out that, since the goal of this paper is to study the effect of ROCOF-based UFLS, the synchronous generators only implement primary frequency control with regulation coefficient of 0.05. Indeed, the effects of secondary frequency control would cover those of any UFLS-scheme. 

\subsection{Wind Farms}
Table~\ref{tab:generation} shows also the nominal capacity and location of the considered wind farms. As it can be noticed, they are installed in 4 different buses, with a total nominal capacity of 1.35 GW. The wind farm plants are assumed to adopt a type-3 double-fed induction generator, that consists in an  asynchronous machine and a back-to-back voltage source converter (VSC). Even though the VSC is simply modeled as an equivalent voltage source, the dynamics resulting from the interaction between control and power system are correctly preserved\cite{1355112}.
For each wind farm, we generate a synthetic power profile, based on real measurements provided %by Electric Reliability Council of Texas
in~\cite{ERCOT}. Since the original data are updated once a minute, we interpolate them over a refined time-scale, whose sampling step is set equal to 1 s. For this analysis, we apply an analytical approach based on iterated smoothing and differencing operations, as described in~\cite{Coughlin-etAl2010}.

\subsection{Load Profiles}
The 19 three-phase dynamic loads are modeled as time-series of active and reactive power absorbed by each load, and thus they are voltage-independent. 
{These power profiles are derived from experimental measurements obtained by PMUs installed on the 125 kV sub-transmission system of the city of Lausanne, Switzerland~\cite{TSGpdc}. Active and reactive power measurements are reported at 50 fps.
The final demand patterns are obtained by rescaling the measured values expressed as per-unit 
to the nominal values of the 39-bus benchmark. Specifically, for each load, every one of the measured per-unit values $x$ is replaced with a scaled value $kx$, $k$ being the nominal load power.
The measured values are obtained from different PMUs during the same time interval.}

\subsection{Phasor Measurement Units}
We equip the network with simulated PMUs, located in each bus containing a load. For this analysis, we adopt P-class PMUs, that implement an enhanced version of interpolated Discrete Fourier Transform (e-IpDFT) to extract synchrophasor, frequency and ROCOF associated to the fundamental component, as further described in~\cite{eIpDFT}. 
The implementation aspects into the Opal-RT RTS is discussed in more detail in~\cite{simulatedPMU}.
The virtual PMUs are characterized by sampling and reporting rate equal to 10 kHz and 50 fps, respectively, and compute ROCOF as the frequency first time-derivative over 20 ms. 
In order to provide a quantitative evaluation of their estimation accuracy, we perform the frequency ramp test as provided by IEEE Std. C.37.118-1. For this analysis, we consider a test signal whose fundamental frequency varies between 45 and 55 Hz with a constant ROCOF of 1 Hz/s. In this context, we define the estimation uncertainty as the standard deviation of ROCOF estimates, equal to 15 mHz/s. Based on these results, P-class PMUs can be employed in the proposed relaying scheme, where the threshold levels differ by almost hundreds mHz/s.  

It is worth pointing out, that the use of simulated PMUs makes the validation of the proposed technique  realistic, as it enables us to evaluate the PMU behaviour during power systems' transients, {with the same metrological performance of the real device described in~\cite{eIpDFT}.}  

\subsection{The Proposed UFLS Relaying Scheme}
\begin{figure}
\centering
\includegraphics[width=0.95\columnwidth]{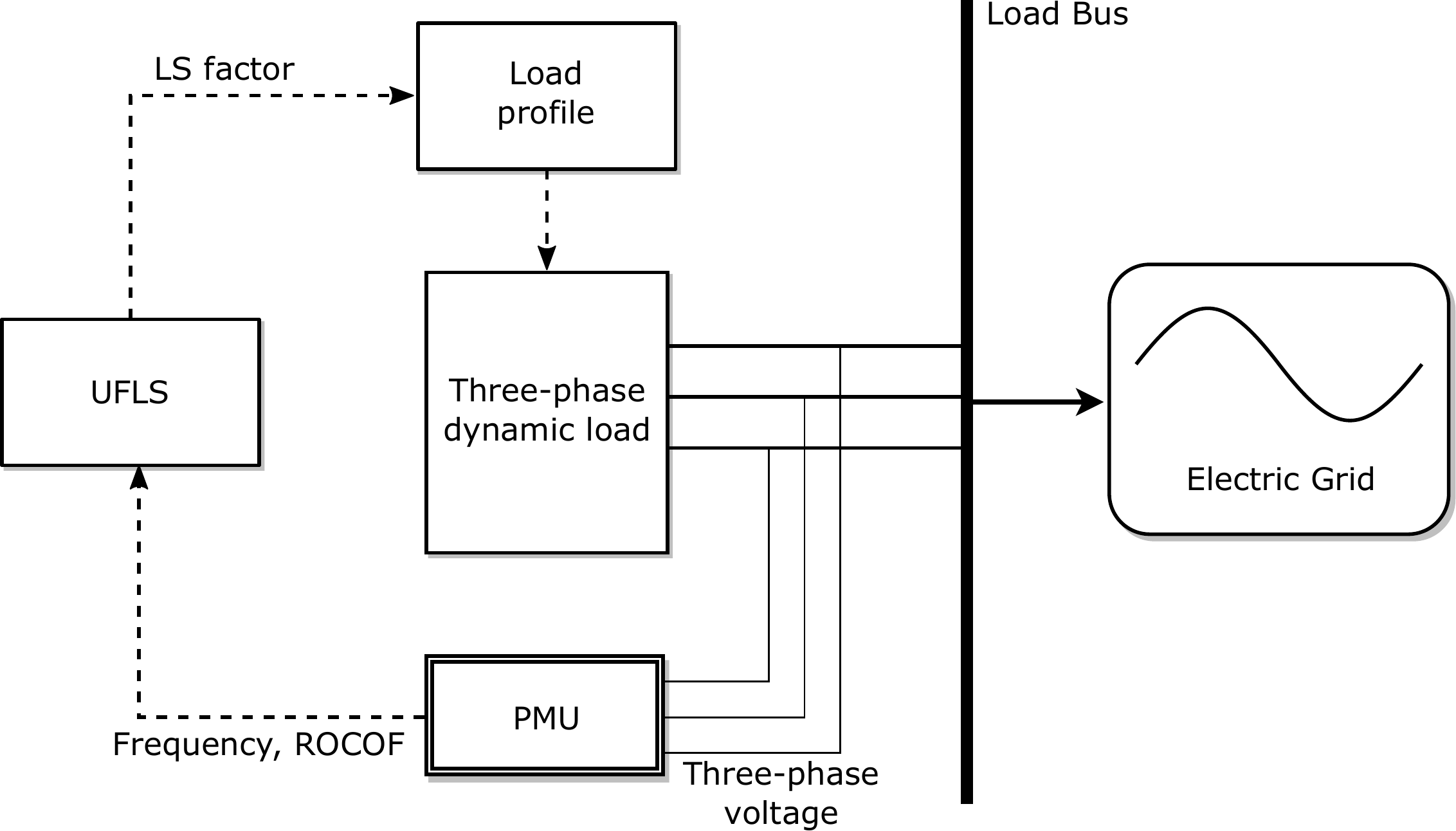}
\caption{The model of the proposed UFLS scheme.}
\label{fig:UFLS_model}
\end{figure}

The proposed UFLS scheme is distributed, in the sense that control actions are taken at each load. As it can be seen in Fig.~\ref{fig:UFLS_model}, 
one PMU is installed at each load bus and measures the bus voltage synchrophasors, as well as frequency and ROCOF. These measurements are sent to the LS controller, that directly acts on the demand profile depending on the adopted UFLS strategy. 
A 500 ms time delay filters ROCOF estimates, whereas a 5 s time delay is implemented between two consecutive restoration actions, to avoid undesirable LS.

\section{Results}
\begin{table}
	\caption{Scenario 1: Nadir Frequency, Maximum LS Factor, Duration of the whole LS and LR Process and Total Curtailed Energy.}
	\label{tab:scen1}
	\centering
	\begin{tabular}{ccccc}
		\toprule
		LS-scheme  & Nadir frequency & Max LS & Duration & Energy \\
		& [Hz]    & [\%]   & [s]         & [MWh] \\
		\midrule
		ROCOF-LS A &  48.72  & 5  & 10.5 & 0.9        \\
		ROCOF-LS B &  49.01  & 25 & 62.7 & 4.8        \\
		\midrule
		f-LS       &  48.77  & 15     & 40.9 & 4.0        \\
		\toprule
	\end{tabular}
\end{table}
\begin{table}
	\caption{Scenario 2: Nadir Frequency, Maximum LS Factor, Duration of the whole LS and LR Process and Total Curtailed Energy.}
	\label{tab:scen2}
	\centering
	\begin{tabular}{ccccc}
		\toprule
		LS-scheme  & Nadir frequency & Max LS & Duration & Energy \\
		& [Hz]    & [\%]   & [s]         & [MWh] \\
		\midrule
		ROCOF-LS A & 48.87 & 35  & 64.5  & 14.8 \\
		ROCOF-LS B & 48.57 & 25  & 63.4  & 15.5 \\
		\midrule
		f-LS       &  48.49  & 25 & 14  & 13.7 \\
		\toprule
	\end{tabular}
\end{table}
To show the impact of different UFLS strategies, we carry out dedicated simulations of two emergency scenarios: 
\begin{itemize}
	\item Scenario 1: G4 and G6 outage, 1 GW tripped power
	\item Scenario 2: G4, G5 and G6 outage, 1.5 GW tripped power
\end{itemize}
The former refers to a weak load imbalance where the frequency would restore after a sufficient amount of time even if no control action is taken, whereas the latter refers to a strong load imbalance, that would lead to a system collapse. 

\begin{figure}
	\centering
	\subfloat[] 
	{\includegraphics[width=0.98\columnwidth]{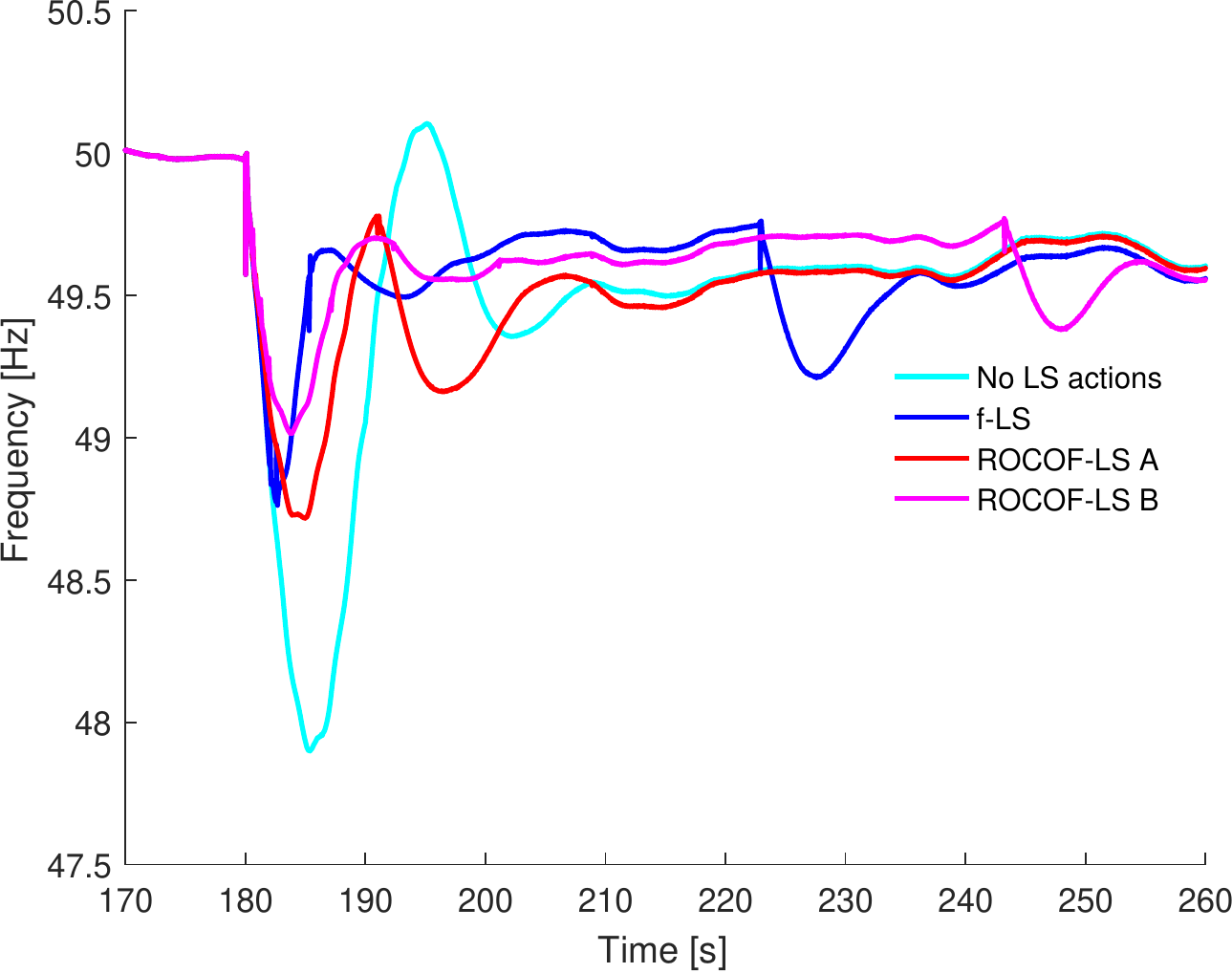} 
		\label{fig:fA}}
	\\
	\subfloat[]
	{\includegraphics[width=0.98\columnwidth]{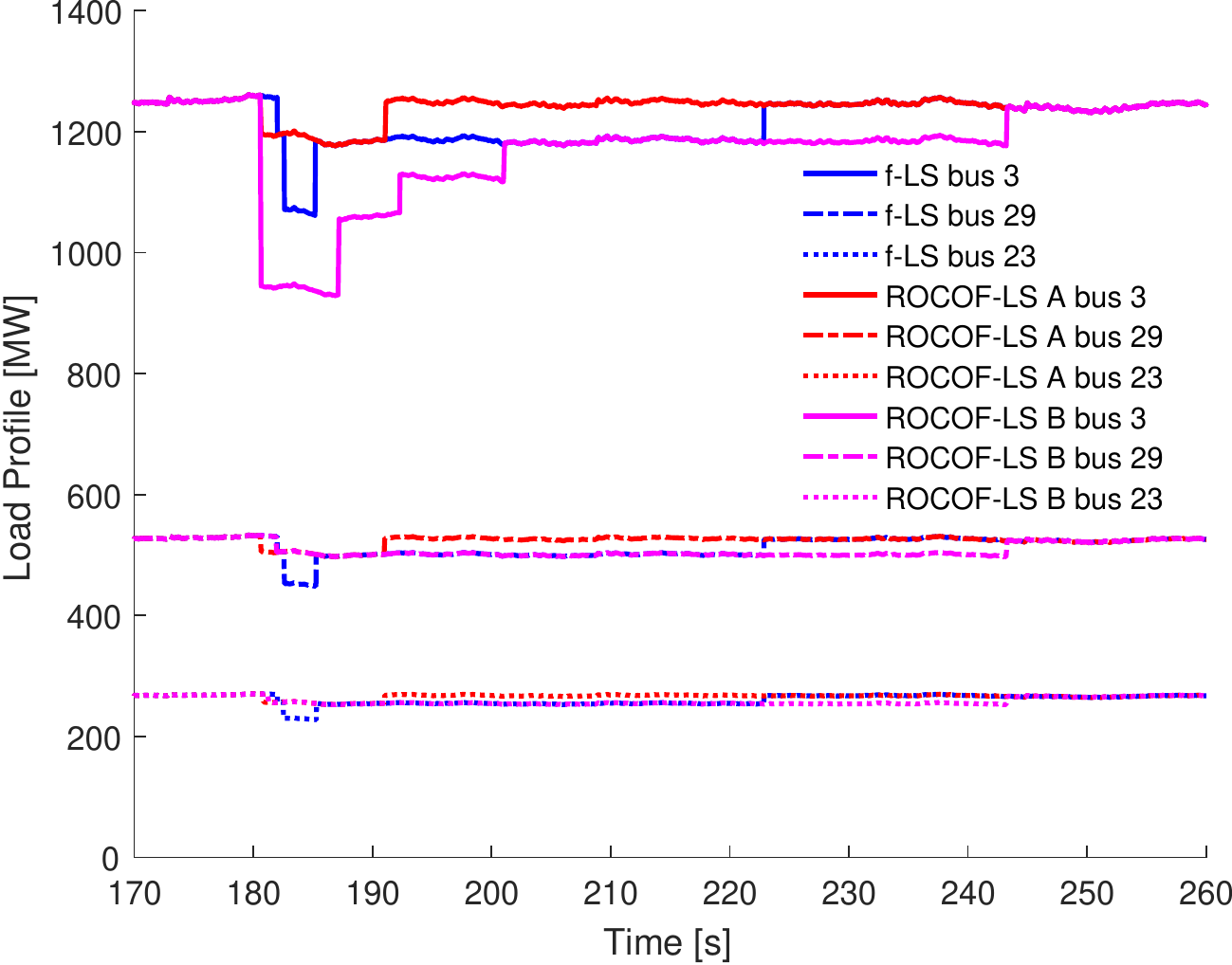}
		\label{fig:lsA}}
	\caption{Scenario 1: (a) frequency at bus 3 and (b) load profiles at bus 3, 29, 23 under different LS-schemes.}
	\label{fig:scen1}
\end{figure}

\begin{figure}
	\centering
	\subfloat[] 
	{\includegraphics[width=0.98\columnwidth]{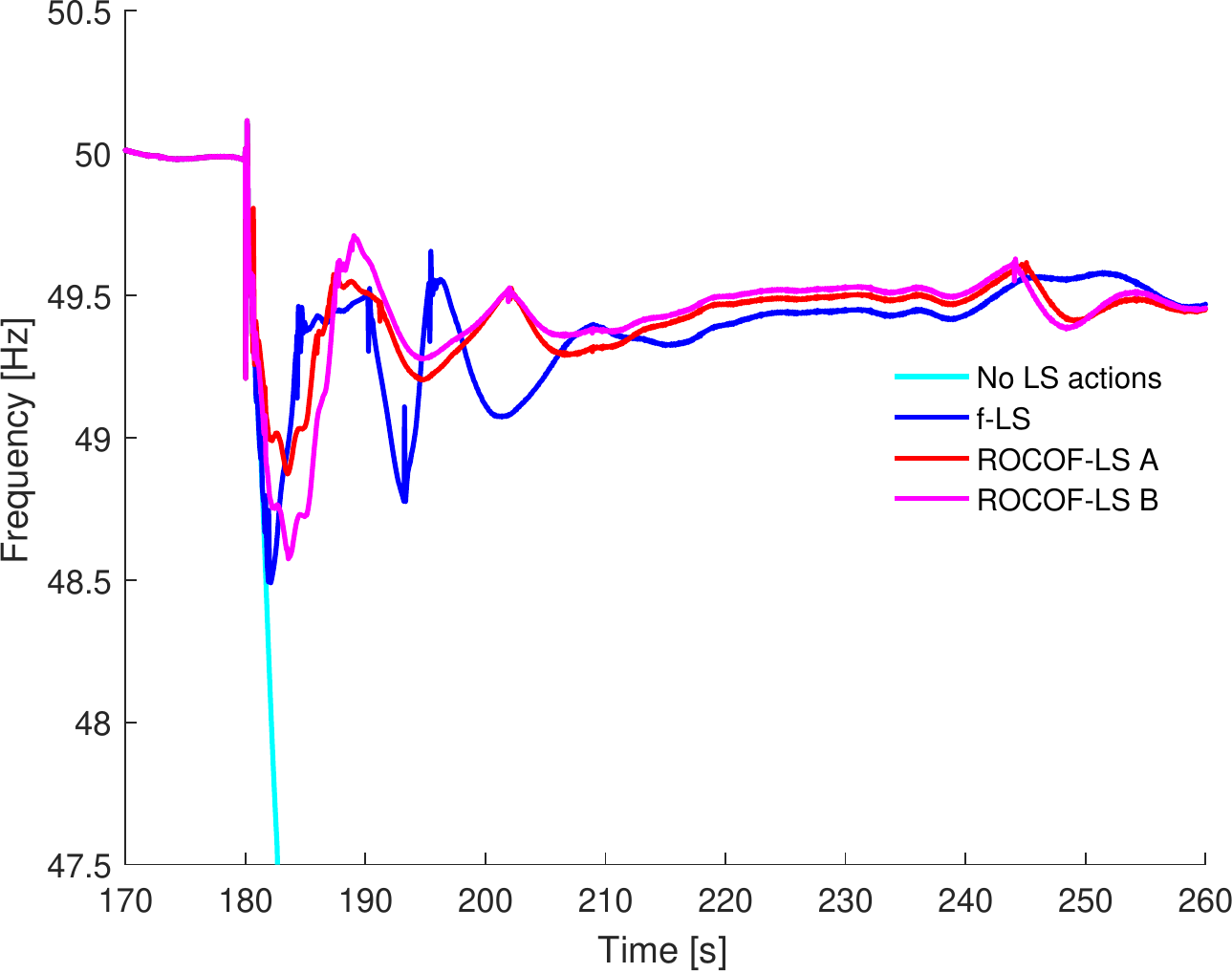} 
		\label{fig:fB}}
	\\
	\subfloat[]
	{\includegraphics[width=0.98\columnwidth]{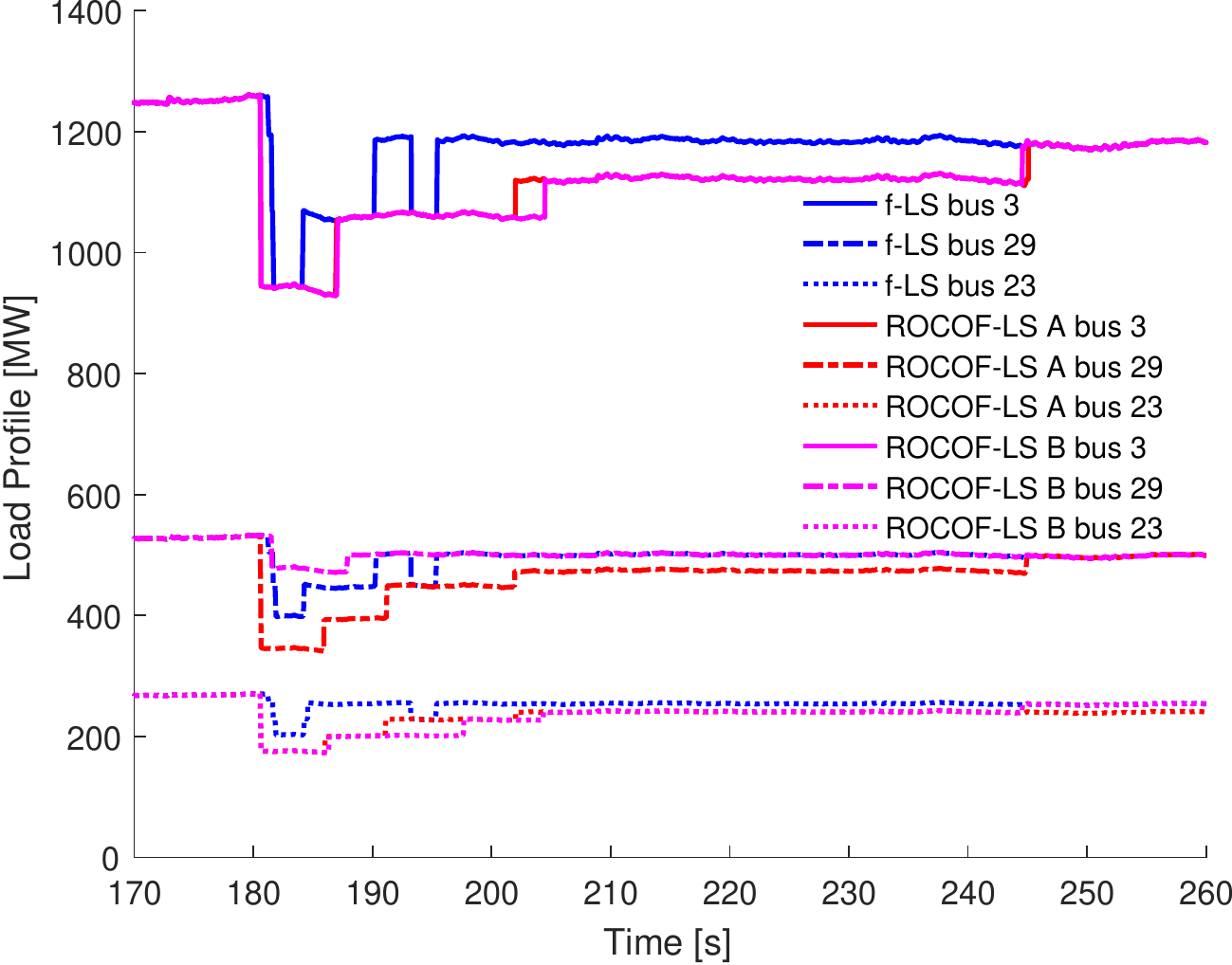}
		\label{fig:lsB}}
	\caption{Scenario 2: (a) frequency at bus 3 and (b) load profiles at bus 3, 29, 23 under different LS-schemes.}
	\label{fig:scen2}
\end{figure}

In this respect, 
Fig.~\ref{fig:scen1}
and~\ref{fig:scen2}
compare the performance of f-LS and ROCOF-LS techniques in terms of frequency and load profiles during scenario 1 and 2, respectively. 
The frequency is measured at bus 4, whereas the loads are reported for three buses (3, 23 and 29, similar trends hold for all buses). 
Table~\ref{tab:scen1}
and~\ref{tab:scen2}
show the obtained nadir frequency, maximum LS factor, duration of the whole LS and LR process as well as total curtailed energy. 

In Scenario 1, if no control action is taken, the system reaches the nadir frequency of 47.9 Hz just 5.3 s after the contingency, with a frequency fall rate of around 0.44 Hz/s.

In case the f-LS scheme is implemented, the frequency exceeds first (48.9 Hz) and second threshold (48.8 Hz) in 2 and 2.5 s, respectively, leading to a percentage of shed loads equal to 5\% and 15\%. Then, frequency increases up to 49.6 Hz in 3 s, restoring loads up to 5\%. Only after 40.9 s, frequency reaches the 49.75-Hz threshold, and it is possible to completely restore the load patterns. The total curtailed energy is 4 MWh.

As regards ROCOF-LS schemes, since estimated ROCOF reaches 0.44 Hz/s as soon as the generators are tripped, the first threshold is exceeded just after 0.6 s. In case A, characterized by higher threshold levels, only the 5\% threshold is exceeded. After an unavoidable drop up to 48.72 Hz, the frequency reaches 49.75 Hz in about 10 seconds and the loads are fully restored. In this way, the total curtailed energy is limited to 0.9 MWh, i.e. a much smaller amount if compared with the f-LS results. In case B, characterized by lower threshold levels, loads are shed more promptly with respect to the f-LS. However, since higher shares of loads are shed (up to 15 \%), the total curtailed energy is comparable or even larger than in the f-LS results. Therefore, the advantages of using ROCOF measurements vanish. 

As it can be seen in Fig.~\ref{fig:fB}, if no control action is taken, Scenario 2 leads to a system collapse, with a frequency fall rate of 0.6 Hz/s. It is worth noticing that none of the tested LR schemes leads to a full restoration, but loads are restored up to 5\%. In fact, since no secondary frequency control is implemented, the frequency cannot be increased over 49.6 Hz.
 
The f-LS scheme produces a percentage of shed loads up to 25\% just 1.9 s after the contingency. The frequency restoration process takes 14 s and involves 13.7 MWh of curtailed energy. 

According to the ROCOF-LS scheme, instead, the estimated ROCOF rapidly exceeds both second and third threshold levels in most buses. This leads to a prompt restoration just 0.6 s after the contingency. Comprehensively, the curtailed energy is equal to 14.8 and 15.5 MWh for case A and B, respectively.

\section{Conclusions}
The paper describes a distributed UFLS and LR scheme, relying on PMU-based measurements of frequency and ROCOF. 
When the frequency decays dramatically revealing a system contingency, ROCOF estimates govern the LS process, whereas frequency estimates govern the LR. 
The performance of the proposed relaying scheme is assessed within a RTS implementing a modified version of the IEEE 39 bus system under 2 different scenarios. 

The results show that under poor system contingencies, a ROCOF-based LS scheme outperforms the frequency-based one, in terms of total curtailed energy (75\% less) and duration (75\% shorter). 
The advantages come from the fact that ROCOF estimates are a faster indicator of system collapse. 
Conversely, under severe contingencies, the performance of the two methods are comparable, because the only way to successfully recover the frequency is to shed a consistent amount of loads. 

Future work will investigate the impact of different synchrophasor estimation algorithms on the proposed relaying scheme performance. Also, a more realistic simulation scenario will take into account the effects of measurement noise.

\bibliographystyle{IEEEtran}
\bibliography{mybibfile}

\end{document}